\documentclass[twocolumn,prl,showpacs,superscriptaddress,a4paper,aps]{revtex4}
\usepackage{epsfig}
\usepackage{dcolumn}
\usepackage{bm}
\begin{document}
\preprint{28-07-2005}

\title{Zero Temperature Hysteresis in Random Field Ising Model on
  Bethe Lattices: approach to mean field behavior with increasing
  coordination number $z$.}
\author{Xavier Illa}
\email{xit@ecm.ub.es}
\affiliation{ Departament d'Estructura i Constituents de la Mat\`eria,
  Universitat de Barcelona \\ Diagonal 647, Facultat de F\'{\i}sica,
  08028 Barcelona, Catalonia}
\author{Prabodh Shukla}
%
%
\affiliation{ Departament d'Estructura i Constituents de la Mat\`eria,
  Universitat de Barcelona \\ Diagonal 647, Facultat de F\'{\i}sica,
  08028 Barcelona, Catalonia}

\affiliation{Physics Department, North Eastern Hill University,
  Shillong-793022, India}
\author{Eduard Vives}
%
%
\affiliation{ Departament d'Estructura i Constituents de la Mat\`eria,
  Universitat de Barcelona \\ Diagonal 647, Facultat de F\'{\i}sica,
  08028 Barcelona, Catalonia}
\date{\today}

\begin{abstract}
  We consider the analytic solution of the zero temperature hysteresis
  in the random field Ising model on a Bethe lattice of coordination
  number $z$, and study how it approaches the mean field solution in
  the limit $z \rightarrow \infty$. New analytical results concerning
  the energy of the system along the hysteresis loop and first order
  reversal curves (FORC diagrams) are also presented.
\end{abstract}


\maketitle

The Random Field Ising model with zero temperature ($T=0$) metastable
dynamics is a prototype lattice model for understanding the dynamics
of disorder driven first-order phase transitions and hysteresis in
several complex systems \cite{Sethna1993}. The model was introduced
more than 10 years ago and studied analytically in the mean-field
limit and numerically on finite dimensional lattices (3d, 4d, etc). It
predicts the existence of a critical point in the systems's response
to a slowly and smoothly varying applied field. Theoretical interest
in the model increased after many of its properties were obtained
analytically on Bethe lattices \cite{Dhar1997}.  It was found that the
critical point exists only on lattices of coordination number $z \ge
4$.  In recent years the analysis on Bethe lattices has been extended
to obtain additional results like the trajectories of the first-order
reversal curves in the field vs. magnetization ($H-m$) diagram
\cite{Shukla2001}, the behavior of the different energy terms in the
hamiltonian \cite{Illa2005} and the demagnetized states
\cite{Alava2005}.  In this Brief Report we present new developments
along three lines: (i) relationship between the mean field solution of
the model and its solution on Bethe lattices of large coordination
numbers, (ii) a different argument for the computation of the energy
terms that allows us to rewrite the existing results for a $z=4$
lattice in a more compact and transparent form valid for an arbitrary
value of $z$, and (iii) the computation of the FORC-diagrams that
allow a compact description of the properties of the First-Order
Reversal Curves (FORC).

The RFIM model in the mean field limit of infinetely weak but
infinitely long range pair interactions is described by the
Hamiltonian,
\begin{equation}
 {\cal H}^{MF}= - \frac{J}{N-1} \sum_{i,j} S_i S_j -\sigma \sum_{i} h_iS_i
-H\sum_{i} S_i \label{mfham} \end{equation} 
Here $J$ is a ferromagnetic exchange interaction of order unity,
$\{S_i=\pm1; i=1,2,\ldots N\}$ denote Ising spins, $\{h_i\}$ are
independent, identically distributed, on-site quenched random fields,
and $H$ is an externally applied uniform field. The quenched fields
are drawn from a Gaussian distribution $\phi(h_i)$ of variance unity
and mean value zero.  The sums over $i$ run over all sites of the
system, and the sum over $j$ runs over the entire $N-1$ spins of the
system that interact with the spin $S_{i}$. The factor $N-1$ dividing
$J$ ensures that the energy of the system is an extensive quantity.

The model on a Bethe lattice (deep interior of an infinite Caley tree
of coordination number $z$) is based on short range interactions. It
is characterized by the Hamiltonian,
\begin{equation} 
{\cal H}^{B}= - \frac{J}{z} \sum_{i,j} S_i S_j -\sigma \sum_{i} 
h_iS_i -H\sum_{i} S_i 
\label{zham}
\end{equation}
The difference between the two look alike Hamiltonians lies in the sum
over $j$. In Eq.~\ref{zham}, the sum over $j$ runs over $z$ nearest
neighbors of each site $i$ ($z=2d$, for a d-dimensional cubic
lattice).  The factor $z$ dividing $J$ in Eq.~\ref{zham} ensures that
the energy remains extensive in the limit $z \rightarrow N$. Although
equations Eqs.~\ref{mfham} and ~\ref{zham} appear to have an identical
form if $z=N-1$, but the topology is different in the two cases.  In
mean field theory every spin interacts with every other spin but this
is not the case on a Bethe lattice.

In both cases (MF and B) the local force on a site can be writen in
terms of a generic local magnetization $m_i$
\begin{equation}
F_i = J m_i + \sigma h_i + H 
\end{equation}
where in the MF case $m_i^{MF}= m^{MF} = \sum_{j=1}^N S_j/N$ (except
for a negligible small correction in the thermodynamic limit) and for
the B case $m_i^B= \sum_{j=1}^z S_j/z$. To fully specify the
metastable $T=0$ dynamics one must fix the initial state (for
instance, $S_i=-1$ at $H=-\infty$) and adiabatically sweep the field
$H$, relaxing the spins according to the rule $S_i=sign(F_i)$.  In the
MF description, since $m_i$ is independent of $i$, the evolution of
the magnetization $m^{MF}(H)$ is trivially determined by the two
coupled equations,
\begin{eqnarray}
m^{MF} & =& 2p^{MF} -1 \label{mf1}\\
p^{MF} & =&  \int_{\frac{-J m^{MF}-H}{\sigma}}^{\infty} 
\phi(h_i) dh_i  \label{mf2}
\end{eqnarray}
where the first equation relates the magnetization with the
probability $p^{MF}$ of finding a spin up and the second one expresses
that probability as an integral of the distribution of random fields .
The above equations \ref{mf1} and \ref {mf2} admit one fixed point
solution for all fields if $\sigma > \sigma_{c} =
\sqrt{\frac{2}{\pi}}J$ and three fixed point solutions (in a certain
field range) below $\sigma_c$. Two of the solutions are stable and one
is unstable. The two stable solutions are obtained by numerical
iteration starting from the initial state $m=-1$ and $m=1$
respectively and correspond to the two halves of the hysteresis loop
in increasing and decreasing applied field. Note that there is no
hysteresis if $\sigma >\sigma_{c}$. The critical point
$(H_c=0,\sigma_c)$ corresponds to a non-equilibrium critical point of
the system \cite{Sethna1993}
\begin{figure}[htb]
\begin{center}
  \epsfig{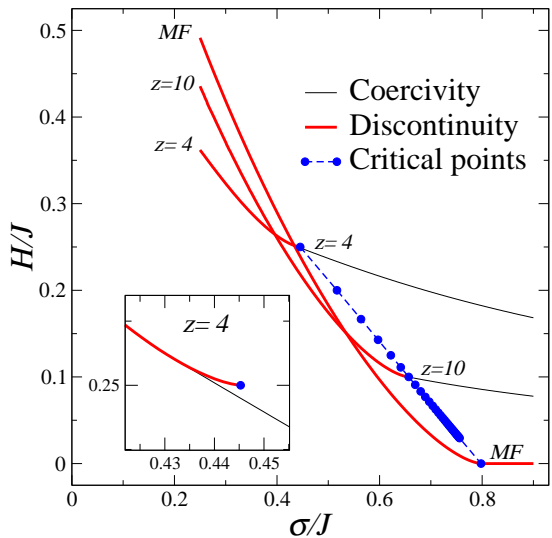}
\end{center}
\caption{\label{FIG1}   
  Phase diagram of the RFIM on bethe lattices as a function of
  increasing $z$, compared with the MF behaviour. The inset shows the
  details around the z=4 critical point. (color online)}
\end{figure}
\begin{figure}[htb]
\begin{center}
  \epsfig{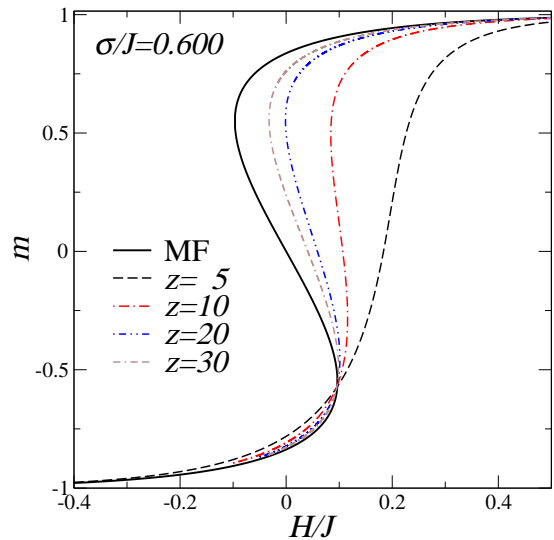}
\end{center}
\caption{\label{FIG2}  
  Magnetization versus field behaviour for $\sigma/J=0.6$ and
  increasing values of $z$ as indicated. The lines are compared with
  the MF behaviour. (color online)}
\end{figure}
The Bethe lattice allows short range fluctuations in the environment
of each site. Therefore $m_i$ is not homogeneous and can take $z+1$
different values ($-1$, $(-z+2)/z$, $\cdots$, $(z-2)/z$, $1$).  Let us
introduce a variable $k=0,\cdots z$ indexing the environments so that
$m_i=(-z+2k)/z $. To find the behaviour of the magnetization $m^B$ (for 
instance, along the $H$-upwards branch of the hysteresis loop) one should 
treat separatelly each of these environments. The
coupled equations become slightly more involved \cite{Dhar1997}:
\begin{eqnarray}
m^B & =& 2p^{B}-1 \\
p^B &=&\sum_{k=0}^{z} {z \choose k} [P^{*}]^{k}[1-P^{*}]^{z-k} P(S_i=+1|k)
\label{PB}
\end{eqnarray} 
where $P(S_i=+1|k)$ is the probability that $S_i=1$ given that
$k$ of its neigbours are up, which is given by \cite{Illa2005}
\begin{equation}
P(S_i=+1|k) = \int_{\frac{-J(2\frac{k}{z}-1)-H}{\sigma}}^{\infty} 
\phi(h_i)dh_i
\end{equation}
and $P^*$ is the solution when $n \rightarrow \infty$ (fixed point) of
the recurrence relation:
\begin{equation} 
P^{(n)}=\sum_{k=0}^{z-1} {z-1 \choose k} [P^{(n-1)}]^{k}[1-P^{(n-1)}]^{z-1-k} 
P(S_i=+1|k)
\label{Pn}
\end{equation}
This means that $P^*$ satisfies:
\begin{equation} 
P^*=\sum_{k=0}^{z-1} {z-1 \choose k} [P^*]^{k}[1-P^*]^{z-1-k} 
P(S_i=+1|k)
\label{Pstar}
\end{equation}
The physical meaning of $P^*$ is the probability that, along the
up-field branch of the hysteresis loop, a spin is $+1$ given that a
neighbour is forced to be down. For the values of $\sigma$ and
$H$ for which it displays multiple fixed points only the stable fixed
point obtained starting from $P^{0}=0$ will have physical meaning. For
small values of $z$ the results of the Bethe lattice are in striking
contrast with those of the MF case due to the neglect of environment
fluctuations in this last case.  The critical point $(H_c,\sigma_c)$
is absent on lattices with $z=2$ (1d model), as well as $z=3$. For
$z\ge 4$, there is a value of the applied field $H \le J/z$ where the
magnetization jumps discontinuously if $\sigma<\sigma_c$. The size of
the jump reduces with increasing $\sigma$ and vanishes as $\sigma$
approaches $\sigma_c$, and $H$ approaches $H_c=J/z$. For
$\sigma>\sigma_c$, there is hysteresis but no discontinuity in the
magnetization. The critical point for $z=4$ is located at $H_c=J/z$
and $\sigma_c\approx.445315J$.  Fig. \ref{FIG1} shows the phase
diagram for different values of $z$.  The critical points for $z=4, 5
\dots 35$ are indicated by filled circles. The discontinuity in the
magnetization occurs on a field $H_{dis}$ (indicated by the thick lines) and 
coercivity ($m=0$) occurs on $H_{coe}$ (thin continuous lines). The inset 
shows a detail of the $z=4$ case, revealing that below $\sigma_c$, 
$H_{coe} \neq H_{dis}$.

Although, the most important differences between the MF and B
approaches are seen for small values of $z$, it is of interest to
study the Bethe lattice hysteresis loops for increasing $z$. In the
limit $z\rightarrow\infty$, we may expect the Bethe lattice results to
approach those of the MF case. The reason is that in this limit the
fluctuations in the exchange field at a site approach zero, and
therefore the model on a Bethe lattice approaches the mean field
model. This argument is not entirely transparent because unlike the
mean field model, the nearest neighbors of a site on a Bethe lattice
are not nearest neighbors of each other for any $z$. However, as we
see from Fig.\ref{FIG1}, the critical point in the limit
$z\rightarrow\infty$ tends to the critical point of the MF theory
$h_c=0,\sigma_c=\sqrt{2/\pi}$.  Indeed, Fig.\ref{FIG2} shows that the
entire magnetization curve tends to the MF result in the limit
$z\rightarrow\infty$. Thus the expectation that the B results fall
over the MF results in this limit is indeed born out by comparing the
two numerically. The equivalence can also be shown analytically from
Eq. \ref{PB} as follows: In the limit $z\rightarrow\infty$,
\begin{eqnarray}
p^B(h)&=&\sum_{k=0}^{z}{z \choose k}[P^{*}]^{k}[1-P^{*}]^{z-k} 
\int_{\frac{-J(2\frac{k}{z}-1)-H}{\sigma}}^{\infty} \phi(h_i)dh_i
\nonumber \\ \Rightarrow p^B(h)&=& \int dk \delta(P^{*}-\frac{k}{z}) 
\int_{\frac{-J(2\frac{k}{z}-1)-H}{\sigma}}^{\infty} \phi(h_i)dh_i
\nonumber \\ \Rightarrow p^B(h) &=& 
\int_{-\left[\frac{J(2P^{*}-1)+h}{\sigma}\right]}^{\infty}
\phi(h_i)dh_i  
\end{eqnarray}
But $(2P^{*}-1)$ $\Rightarrow$ $(2p^B-1)=m^B$ in the limit
$z\rightarrow\infty$. Thus the equation determining the magnetization
in an applied field $H$ becomes the same in the two strategies.
\begin{figure}[htb]
\begin{center}
  \epsfig{file=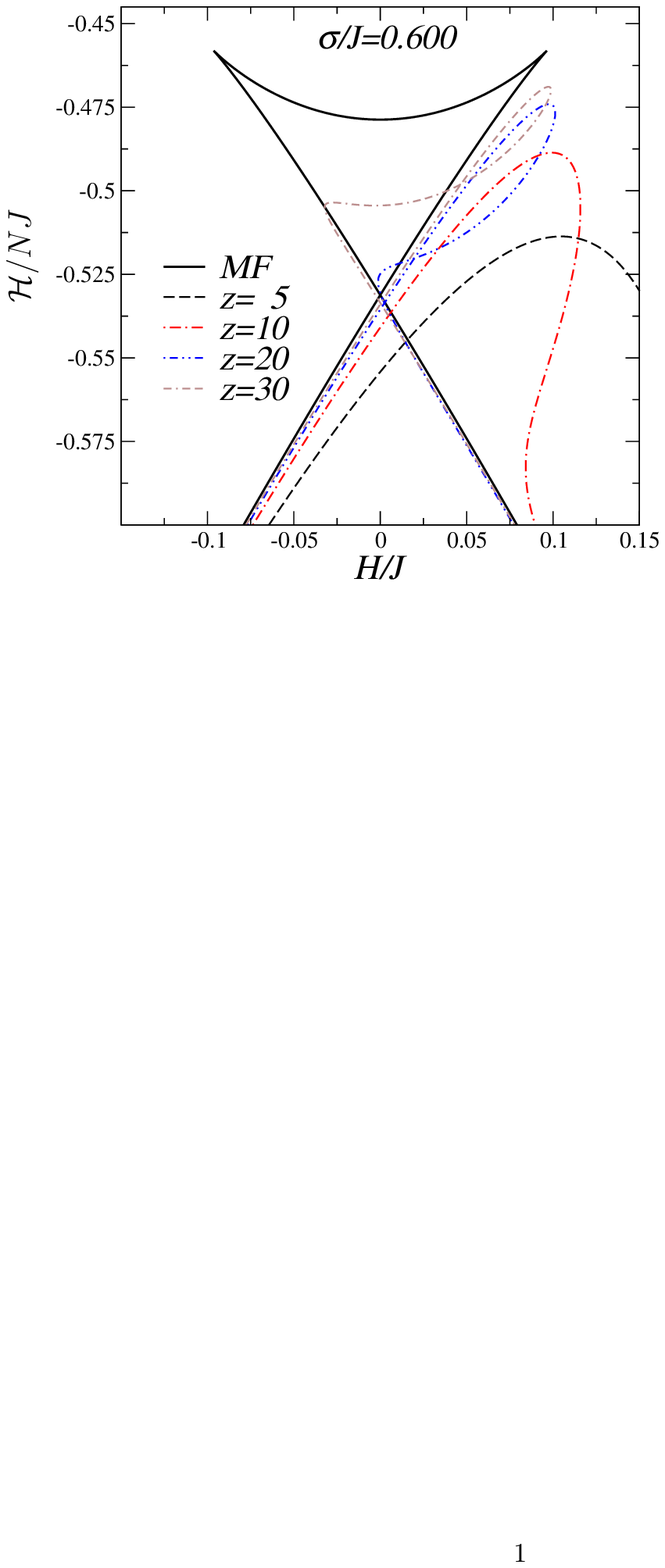,width=7.5cm,clip= }
\end{center}
\caption{\label{FIG3}  Behaviour of the reduced hamiltonian ${\cal H}^B/JN$ 
  as a function of $H/J$ for $\sigma/J=0.6$ and increasing values of
  $z$, compared with the MF behaviour. (color online)}
\end{figure}

\begin{figure*}[htb]
\begin{center}
  \epsfig{file=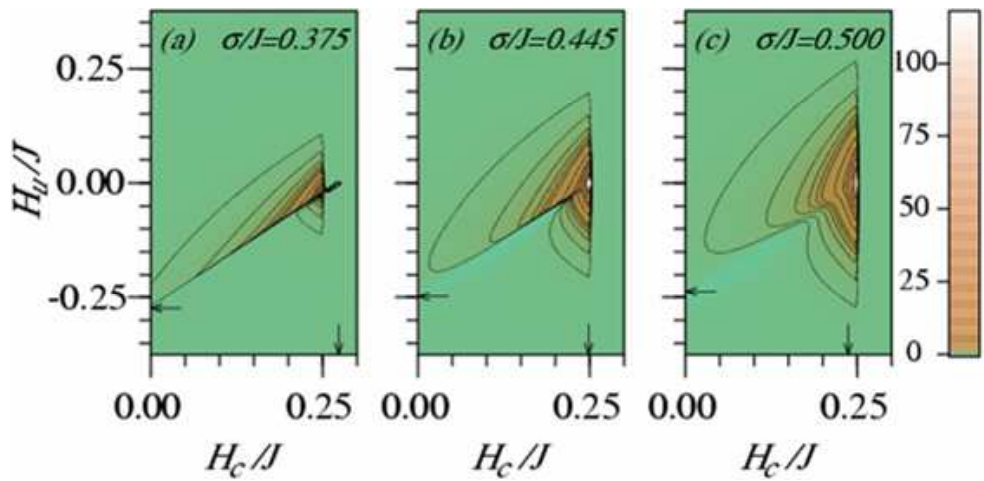,width=13.0cm,clip= }
\end{center}
\caption{\label{FIG4}  First-order reversal curves and FORC diagram for z=4 
  and three different values of $\sigma/J$, as indicated. Arrows indicate 
the value of the coercive field $H_{coe}/J$ on the two axes. (color online)}
\end{figure*}

Let us now focus our attention of the exchange interaction along the
hysteresis loop on the Bethe lattice. In a previous calculation
\cite{Illa2005} the exchange interaction was computed from the two
site probability $p(S_i,S_j)$. Here we will show that it can be
written as a single site equation. Let us start from Eq. (11) in Ref.
\onlinecite{Illa2005} that gives the correlation of two neigbouring
sites: $\langle S_i S_j \rangle$. By introducing the results of Eq.
(16), (21), (22) and (23) of the same reference one can write:
\begin{equation}
\langle S_i S_j \rangle = 1 - 4 P^* + 4 P^* Q^*
\label{corr}
\end{equation}
where $Q^*$ is given by:
\begin{equation}
Q^*=\sum_{k=0}^{z-1} {z-1 \choose k} \left [ P^* \right ]^k 
\left [1-P^* \right ]^{z-1-k} P(S_i=+1|k+1)
\label{qstar}
\end{equation}
Similarly to $P^*$, $Q^*$ can be read as the probability that (along
the increasing field branch) a spin is $+1$ given that a neigbour is
forced to be up.Equation (\ref{corr}) is still a two site equation since its last
term in (\ref{corr}) contains the double sum with two indexes
reflecting the state of the $z-1$ spins (different from $S_i$ and
$S_j$) in the neigbourhoods of $S_i$ and $S_j$. Now we make use of the
identity:
\begin{equation}
1= \sum_{k=0}^{z} {z \choose k} \left [ P^* \right ]^k 
\left [1-P^* \right ]^{z-k}
\end{equation}
Taking the derivative with respect to $P^*$ one gets
\begin{equation}
P^* = \sum_{k=0}^{z} {z \choose k} \left[ P^{*k} \right ] 
\left[1-P^*\right ]  ^{z-k} \frac{k}{z}
\label{pmod}
\end{equation}
From Eq. \ref{qstar}, rearranging the indices ($j=k+1$) it is easy to
obtain:
\begin{equation}
P^* Q^* = \sum_{k=0}^{z} {z \choose k} \left [  P^*\right]^{k}   \left[1-P^*\right]^{z-k} \frac{k}{z}  
P(S_i=+1|k)
\label{pq}
\end{equation}

Introducing (\ref{pmod}) and (\ref{pq}) in (\ref{corr}), and taking
into account that $\langle U_e \rangle / N = -\frac{J}{2} \langle S_i
S_j \rangle$ one gets: \footnote{Note that, given the different
  definition of the Hamiltonian $H^B$ in the present paper, there is a
  missing $z$ factor in $U_e$ when comparing with Ref.\cite{Illa2005}}
\begin{eqnarray}
& & \langle U_e  \rangle /N =  -\frac{1}{2}J + \\  
& & + \sum_{k=0}^{z} {z \choose k}
[P^{*}]^{k}[1-P^{*}]^{z-k} 
\frac{2Jk}{z} \left [1-P(S_i=+1|k) \right ]  \nonumber
\end{eqnarray}
This equation shows that the exchange energy can be computed as the
energy corresponding to the saturated state plus an exces
energy ($2J/z$) associated to each broken bond. This computation also
allows to write the full average hamiltonian $\cal H$ as a constant
term plus a sum over the state of a single site environment. Figure
\ref{FIG3} shows the behaviour of ${\cal H}/J N$ as a function of the
external field $H/J$ for $\sigma/J=0.6$ and increasing values of $z$
as indicated. As can be seen the behaviour also tends to the MF
behaviour shown by a dashed line. It is interesting to note that even
the unphysical states in the MF curve are recovered from the limit of
the unphysical states of the Bethe lattice with finite $z$.

The third analysis that we want to present is that of the FORC-diagrams. 
Such diagrams were introduced \cite{Pike1999} in order to
simplify the description of the collection of first-order reversal
curves which describe the magnetization $m(H_2,H_1)$ obtained starting
from saturation, adiabatically decreasing the field until $H_1$ and
subsequently increasing the field up to $H_2$. The FORC-diagrams are
computed by evaluating $\rho=\partial^2 m / \partial H_1 \partial
H_2$. This second derivative is represented as a function of
$H_u=(H_2+H_1)/2 $ and $H_c=(H_2-H_1)/2$.  The fact that FORC can be
computed exactly on Bethe lattices \cite{Shukla2001},
which for $z \ge 4$ exhibit a disorder induced phase transition, allows
a better understanding of some interesting features of the FORC-diagrams. 

Fig. \ref{FIG4} exhibits the FORC-diagrams corresponding to $z=4$.
They have been numerically computed by evaluating the function $\rho$
in steps of $\Delta H_1=\Delta H_2 = 0.01 J/z $ for
three values of $\sigma$ as indicated. The first interesting feature, which is
clearly seen in Fig. \ref{FIG4}(c) is the existence of a well defined
peak for $H_u\simeq 0$ and $H_c\simeq H_{coe}\simeq J/z$.  This
indicates that the maximum variation in the slope $\partial m/\partial
H_2$ occurs around $H_1=-H_{coe}$ and $H_2=H_{coe}$. The
second remarkable property is the fact that $\rho=0$ almost everywhere
for $H_c>J/z$. (The only exception is the infinitessimally thin ``ridge''
entering such a region for $\sigma<\sigma_c$). The reason is that on
the Bethe lattice, FORC become independent of $H_2$ when $H_2-H_1>2J/z$
since, as pointed in Ref. \cite{Shukla2001}, the solutions
joint (and merge) the main
hysteresis loop. The third feature to notice in Fig.
\ref{FIG4} is the ``valley'' along an approximate line $H_c-H_u \sim J/z$.
The reason for this valley is that the slope of the FORC
$\partial m/\partial H_2$ increases with decreasing $H_1$, is maximum
when the reversing field $H_1$ is close to $-H_{coe}$, and 
decreases again for more negative reversing fields ($H_1<-H_{coe}$). 
This valley is smooth for large $\sigma$ but becomes sharper when $\sigma
\rightarrow \sigma_c$. Below $\sigma_c$ it transforms into a
discontinuity ``cliff'', due to the occurence of the discontinuity in the
hysteresis loop. The discontinuity in the loop also explains the
existence of the infinitessimally thin ridge, when $\sigma<\sigma_c$
and $J/z<H_c<H_{dis}$.

P.S. acknowledges the hospitality of the Universitat
de Barcelona and support from Generalitat de
Catalunya (Project 2004PIV2-0002) . X.I. acknowledges a grant from
DGI-MEC (Spain). We also acknowledge finantial support from projects
MAT2004-01291 (CICyT, Spain) and SGR-2001-00066 (Generalitat de
Catalunya).

\vspace{-1mm}


\end{document}